\documentclass[journal=jacsat]{achemso}

\author{Natalia A. Denesyuk}
\author{D. Thirumalai}
\email{thirum@umd.edu}
\affiliation{Department of Chemistry and Biochemistry and Biophysics Program, Institute for Physical Science and Technology,
University of Maryland, College Park, Maryland 20742}
\title{A Coarse-Grained Model for Predicting RNA Folding Thermodynamics}
\keywords{Hairpin, pseudoknot, stability, ionic strength}

\renewcommand{\baselinestretch}{2}

\begin{document}
\vspace{-11pt}
\centerline{Tel.: 301-405-4803}
\centerline{Fax: 301-314-9404}

\clearpage

\begin{abstract}
We present a thermodynamically robust coarse-grained model to simulate folding of RNA in monovalent salt solutions. 
The model includes stacking, hydrogen bond and electrostatic interactions as fundamental components in describing the  
stability of RNA structures. The stacking interactions are parametrized using a set of nucleotide-specific parameters, 
which were calibrated against the thermodynamic measurements for single-base stacks and base-pair stacks. All hydrogen 
bonds are assumed to have the same strength, regardless of their context in the RNA structure. The ionic buffer is modeled 
implicitly, using the concept of counterion condensation and the Debye-H{\"u}ckel theory. The three 
adjustable parameters in the model were determined by fitting the experimental data for two RNA hairpins and a pseudoknot.
A single set of parameters provides good agreement with thermodynamic data for the three RNA molecules over a 
wide range of temperatures and salt concentrations. In the process of calibrating the model, we establish the extent of 
counterion condensation onto the single-stranded RNA backbone. The reduced backbone charge is independent of the ionic 
strength and is 60\% of the RNA bare charge at 37 $^{\circ}$C. Our model can be used to predict the folding thermodynamics 
for any RNA molecule in the presence of monovalent ions. 
\end{abstract}

Keywords: Hairpin, pseudoknot, melting temperature, ionic buffer

\clearpage

\section{Introduction}

Since the landmark discovery that RNA molecules can act as enzymes,~\cite{DoudnaNature02} an increasing repertoire of 
cellular functions has been associated with RNA, raising the need to understand how these complex molecules fold into 
elaborate tertiary structures. In response to this challenge, great strides have been made in describing RNA 
folding\cite{Treiber99COSB,Thirumalai05Biochem,Chen08ARBiophys,Woodson11ACR}. Single molecule and ensemble experiments 
using a variety of biophysical methods, combined with theoretical techniques, have led to a conceptual framework for 
interpreting the thermodynamics and kinetics of RNA 
folding~\cite{Zhuang00Science,Russell01JMB,Thirumalai96ACR,Tinoco99JMB,Woodson05COCB}. Despite these advances, there are 
very few reliable structural models with the ability to quantitatively predict the thermodynamic properties of RNA
(see, however, 
refs.~11--17\nocite{Tan10BJ,Hyeon05PNAS,Tan11BJ,Kirmizialtin12BJ,Tan08BJ,Ha03Macromolecules,LuceySchulten12JPCB}). 
The development of simple and accurate models is complicated by the interplay of several energy and length scales, which 
arise from stacking, hydrogen bond and electrostatic interactions. Although multiple interactions contribute to the 
stability of RNA, the most vexing of  these are the electrostatic interactions, since the negatively charged phosphate 
groups  make RNA a strongly charged  polyelectrolyte.~\cite{Thirumalai01ARPC} Because of the strong intramolecular Coulomb 
repulsion, the magnitude of the charge on the phosphate groups has to be reduced in order for RNA molecules to fold. The 
softening of repulsion between the phosphate groups requires the presence of counterions. A number of factors such as the 
Debye length, the Bjerrum length, the number of nucleotides in RNA, as well as the size, valence and shape of 
counterions~\cite{Koculi06JMB} modulate electrostatic interactions, which  further complicates the prediction of RNA 
folding thermodynamics.  

In principle, all-atom simulations of RNA in water provide a straightforward route to computing RNA folding thermodynamics. 
However, uncertainties in nucleic acid force fields and the difficulty in obtaining adequate conformational sampling have 
prevented routine use of all-atom simulations to study the folding of even small RNA molecules. At the same time, the 
success of using polyelectrolyte theories\cite{Thirumalai01ARPC} and simulations\cite{Koculi07JACS} in capturing many salient features of RNA folding justifies the
development of coarse-grained (CG) models. None of the existing CG models of RNA, which have been remarkably successful in 
a variety of applications,~\cite{Whitford09BJ,Cho09PNAS,Lin08JACS,Feng11JACS,Denesyuk11JACS} have been used to reproduce folding thermodynamics over a wide range of ion concentrations and 
temperature. In this paper, we introduce a force field based on a CG model in which each nucleotide is 
represented by three interactions sites (TIS) --- a phosphate, a sugar and a base\cite{Hyeon05PNAS}. The TIS force field includes stacking, 
hydrogen bond and electrostatic interactions that are known to contribute significantly to the stability of RNA structures.
We obtain the thermodynamic parameters for the stacking and hydrogen bond interactions by matching the simulation and 
experimental melting data for various nucleotide dimers and for the pseudoknot from mouse mammary tumor virus (MMTV PK 
in Figure \plainref{SS}). Our description of the electrostatic interactions in RNA relies on the concept of counterion condensation, 
which posits that counterions condense onto the sugar-phosphate backbone and partially reduce the charge on each phosphate 
group. Our simulations provide a way to determine the magnitude of the reduced backbone charge by fitting the experimental 
data for the ion-dependent stability of RNA hairpins (L8 and L10 in Figure \plainref{SS}). Remarkably, experimental data on folding 
thermodynamics of the  MMTV PK, L8,  and L10 are reproduced well over a wide range of temperatures and concentrations of 
monovalent salt using a single set of force field parameters. Our CG force field is transferable, and hence can be adopted 
for other RNA molecules as well.

\section {Methods}

\subsection{Three Interaction Site (TIS) Representation of RNA}

In the TIS model, each nucleotide is replaced by three spherical beads P, S and B, representing a phosphate, 
a sugar and a base (Figure \plainref{STACK}). The coarse-grained beads are at the center of mass of the chemical groups. The energy
function in the TIS model, $U_{\rm TIS}$, has the following six components,
\begin{equation}
U_{\rm TIS}=U_{\rm BL}+U_{\rm BA}+U_{\rm EV}+U_{\rm ST}+U_{\rm HB}+U_{\rm EL},
\label{TIS}
\end{equation}
which correspond to bond length and angle constraints, excluded volume repulsions, single strand base stacking, 
inter-strand hydrogen bonding and electrostatic interactions. We constrain bond lengths, $\rho$, and angles, 
$\alpha$, by harmonic potentials, $U_{\rm BL}(\rho)=k_{\rho}(\rho-\rho_0)^2$ and 
$U_{\rm BA}(\alpha)=k_{\alpha}(\alpha-\alpha_0)^2$, where the equilibrium values $\rho_0$ and $\alpha_0$ are obtained by 
coarse-graining an ideal A-form RNA helix\cite{IdealRNAsite}. The values of $k_{\rho}$, in units of 
kcal$\,$mol$^{-1}$\r{A}$^{-2}$, are: 64 for an S$\rightarrow$P bond, 23 for an P$\rightarrow$S bond (``$\rightarrow$'' 
indicates the downstream direction) and 10 for an S$-$B bond. The values of $k_{\alpha}$ are 
5 kcal$\,$mol$^{-1}$rad$^{-2}$ if the angle involves a base, and 20 kcal$\,$mol$^{-1}$rad$^{-2}$ otherwise. 

Excluded volume between the interacting sites is modeled by a Weeks-Chandler-Andersen (WCA) potential,
\begin{eqnarray}
U_{\rm EV}(r)&=&\varepsilon_0\left[\left(\frac {D_0}{r}\right)^{12}-2\left(\frac {D_0}{r}\right)^{6} + 1\right],\ r\le D_0, \nonumber \\
U_{\rm EV}(r)&=&0,\ r > D_0, 
\label{POT2}
\end{eqnarray}
which has been commonly used to study excluded volume effects in fluids.~\cite{Chandler83Science} The precise form of 
$U_{\rm EV}(r)$ will not affect the results as long as $U_{\rm EV}(r)$ is short-ranged. The WCA potential is 
computationally efficient because it vanishes exactly beyond the contact distance $D_0$. To allow close approach between 
two bases that stack flat one on top of another, we assume $D_0=3.2$ \r{A} and $\varepsilon_0=1$ kcal/mol for the 
interacting sites representing bases. With the exception of stacked bases, this choice of parameters underestimates the 
distance of closest approach between coarse-grained RNA groups. However, to keep the parameterization of the model as 
simple as possible, we use the same $D_0$ and $\varepsilon$ for all interacting sites. We note that the specific choice of 
parameters in Eq.~(\plainref{POT2}) has little effect on the results obtained. In our simulations, stable folds are 
sustained by stacking and hydrogen bond interactions, $U_{\rm ST}$ and $U_{\rm HB}$, which are parameterized using 
experimental thermodynamic data and accurate approach distances between various RNA groups (see below).

\subsection {Stacking Interactions}

Single strand stacking interactions, $U_{\rm ST}$, are applied to any two consecutive nucleotides along the chain,
\begin{equation}
U_{\rm ST}=\frac{U_{\rm ST}^0}{1+k_r(r-r_0)^2+k_{\phi}(\phi_1-\phi_{1,0})^2+k_{\phi}(\phi_2-\phi_{2,0})^2},
\label{UST}
\end{equation}
where $r$, $\phi_1$ and $\phi_2$ are defined in Figure \plainref{STACK}. Sixteen distinct nucleotide dimers are modeled with 
different $r_0$, $\phi_{1,0}$, $\phi_{2,0}$ and $U_{\rm ST}^0$. The structural parameters $r_0$, $\phi_{1,0}$ and 
$\phi_{2,0}$ are obtained by coarse-graining an A-form RNA helix.~\cite{IdealRNAsite} To estimate standard deviations of 
$r$ and $\phi_1$, $\phi_2$ from the corresponding values in a A-form helix, we used double helices in the NMR structure of 
the pseudoknot from human telomerase RNA~\cite{Theimer05MolCell} (PDB code 2K96). We chose this pseudoknot because it has 
two fairly long stems containing six and nine Watson-Crick base pairs. We had previously conducted simulations of the two 
stems at $15\ ^{\circ}$C in the limit of high ionic strength~\cite{Denesyuk11JACS} and found that, for $k_r=1.4$ \r{A}$^{-2}$ and 
$k_{\phi}=4$ rad$^{-2}$, the time averages of $(r-r_0)^2$, $(\phi_1-\phi_{1,0})^2$ and $(\phi_2-\phi_{2,0})^2$ agreed well 
with the standard deviations computed from the NMR structure. The time averages were not very sensitive to a specific 
choice of $U_{\rm ST}^0$. Using $k_r=1.4$ \r{A}$^{-2}$ and $k_{\phi}=4$ rad$^{-2}$, we derive $U_{\rm ST}^0$ from 
available thermodynamic measurements of single-stranded and double-stranded 
RNA,~\cite{Xia98Biochem,Bloomfield00Book,Dima05JMB} as described below.

\subsubsection {Thermodynamic parameters of dimers from experiments}

In the nearest neighbor model of RNA duplexes, the total stability of a duplex is given by a sum of successive 
contributions $\Delta G{x-y\choose w-z}$, where $x-y$ denotes a base pair stacked over the preceding base pair $w-z$. 
The enthalpy, $\Delta H$, and entropy, $\Delta S$, components of $\Delta G{x-y\choose w-z}$ are known 
experimentally at 1 M salt concentration\cite{Xia98Biochem}. Here, we make the following assumptions:
\begin{eqnarray}
\Delta H{x-y\choose w-z}&=&\Delta H{x\choose w} + \Delta H{z\choose y} + 0.5\Delta H(w-z) + 0.5\Delta H(x-y), \nonumber\\
\Delta S{x-y\choose w-z}&=&\Delta S{x\choose w} + \Delta S{z\choose y}, 
\label{dG2}
\end{eqnarray}
where $\Delta H{x\choose w}$ and $\Delta S{x\choose w}$ are the thermodynamic parameters associated with stacking of 
$x$ over $w$ along $5'\rightarrow3'$ in one strand. Additional enthalpy gain $\Delta H(w-z)$ arises from 
hydrogen bonding between $w$ and $z$ in complementary strands. Our goal is to solve Eqs.~(\plainref{dG2}) for 
$\Delta H{x\choose w}$, $\Delta S{x\choose w}$ and $\Delta H(w-z)$. Since the number of unknowns exceeds the number
of equations, we have to make some additional assumptions.

We average the thermodynamic parameters on the left-hand side of Eqs.~(\plainref{dG2}) for stacks 
${{\rm U-A}\choose{\rm A-U}}$ and ${{\rm A-U}\choose{\rm U-A}}$, 
${{\rm A-U}\choose{\rm C-G}}$ and ${{\rm U-A}\choose{\rm G-C}}$, and ${{\rm U-A}\choose{\rm C-G}}$ and 
${{\rm A-U}\choose{\rm G-C}}$, because these values are similar within experimental uncertainty.\cite{Xia98Biochem}
This allows us to assign $\Delta H{x\choose w}=\Delta H{w\choose x}$ and $\Delta S{x\choose w}=\Delta S{w\choose x}$ on the 
right-hand side of Eqs.~(\plainref{dG2}) for all dimers, except for ${{\rm C}\choose{\rm G}}$ and 
${{\rm G}\choose{\rm C}}$. Additional simplifications result from the analysis of experimental data on stacking of 
nucleotide dimers.\cite{Bloomfield00Book,Florian99JPCB} Experiments indicate that dimers ${{\rm A}\choose{\rm A}}$, 
${{\rm U}\choose{\rm A}}$ and ${{\rm C}\choose{\rm A}}$ have similar stacking propensities and can
therefore be described by one set of thermodynamic parameters. The same holds for ${{\rm C}\choose{\rm C}}$ and 
${{\rm U}\choose{\rm C}}$.

The melting temperature of dimer ${{\rm A}\choose{\rm A}}$ is known from experiment, 
$T_{\rm m}=26$ $^{\circ}$C~\cite{Bloomfield00Book}. According to the assumptions above, dimer ${{\rm U}\choose{\rm A}}$ has 
the same melting temperature. Combining Eqs.~(\plainref{dG2}) for ${{\rm U-A}\choose{\rm A-U}}$ and the relationship 
$\Delta H{{\rm U}\choose{\rm A}}=k_{\rm B}T_{\rm m}\Delta S{{\rm U}\choose{\rm A}}$, where $T_{\rm m}=299$ K and 
$k_{\rm B}$ is the Boltzmann constant, we can solve for $\Delta H{{\rm U}\choose{\rm A}}$, 
$\Delta S{{\rm U}\choose{\rm A}}$ and $\Delta H({\rm A}-{\rm U})$. 
By assigning $\Delta H{{\rm A}\choose{\rm A}}=\Delta H{{\rm U}\choose{\rm A}}$ and
$\Delta S{{\rm A}\choose{\rm A}}=\Delta S{{\rm U}\choose{\rm A}}$ in Eqs.~(\plainref{dG2}) for 
${{\rm A-U}\choose{\rm A-U}}$, we solve for $\Delta H{{\rm U}\choose{\rm U}}$ and 
$\Delta S{{\rm U}\choose{\rm U}}$. Finally, we assume
\begin{eqnarray}
\Delta H{{\rm U}\choose{\rm C}}=k\Delta H{{\rm U}\choose{\rm A}} + (1-k)\Delta H{{\rm U}\choose{\rm U}},\nonumber\\
\Delta S{{\rm U}\choose{\rm C}}=k\Delta S{{\rm U}\choose{\rm A}} + (1-k)\Delta S{{\rm U}\choose{\rm U}},
\label{CU}
\end{eqnarray}
where $k$ is a constant. This assumption is based on the observation that the measured enthalpy changes of duplex 
formation, $\Delta H{x-y\choose w-z}$ in Eqs.~(\plainref{dG2}), are approximately in proportion to the corresponding 
entropy changes, $\Delta S{x-y\choose w-z}$. Furthermore, from previous assumptions, the melting temperature of 
dimer ${{\rm U}\choose{\rm C}}$ should match the $T_{\rm m}$ of ${{\rm C}\choose{\rm C}}$, which is known experimentally to be 
13 $^{\circ}$C\cite{Bloomfield00Book}. Using this result and Eqs.~(\plainref{CU}), we obtain 
$\Delta H{{\rm U}\choose{\rm C}}$ and $\Delta S{{\rm U}\choose{\rm C}}$.

The enthalpies of hydrogen bond formation between Watson-Crick base pairs are related as 
$\Delta H({\rm G}-{\rm C})=3/2 \Delta H({\rm A}-{\rm U})$, where $\Delta H({\rm A}-{\rm U})=-1.47$ kcal/mol is the result
of the calculation outlined above. The remaining thermodynamic parameters follow directly from Eqs.~(\plainref{dG2}) 
with no further approximations. The results are summarized in Table \plainref{tbl:table1}. The relative stacking 
propensities of dimers in Table \plainref{tbl:table1} are consistent with experiments\cite{Bloomfield00Book,Florian99JPCB}.

\subsubsection {Thermodynamic parameters of dimers from simulations}

To calibrate the model, we simulated stacking of coarse-grained dimers similar to that shown in 
Figure \plainref{STACK}. We used the stacking potential $U_{\rm ST}$ in Eq.~(\plainref{UST}) with 
$U_{\rm ST}^0=-h+k_{\rm  B}(T-T_{\rm  m})s$, where $T$ (K) is the 
temperature, $T_{\rm  m}$ (K) is the melting temperature given in Table \plainref{tbl:table1}, and $h$ and $s$ are adjustable 
parameters. In simulations, we computed the stability $\Delta G$ of stacked dimers at temperature $T$ using
\begin{equation}
\Delta G = -k_{\rm  B}T\ln{p} + k_{\rm  B}T\ln{(1-p)} + \Delta G_0, 
\label{eqn:dGsim}
\end{equation}
where $p$ is the fraction of all sampled configurations for which $U_{\rm{ST}}<-k_{\rm B}T$ (Figure \plainref{fgr:UST1}). 
The correction $\Delta G_0$ in Eq.~(\plainref{eqn:dGsim}) is assumed to be constant for all dimers and accounts for any 
differences in the definition of $\Delta G$ between experiments and simulations.  

Figure \plainref{fgr:UST1} shows the simulation values of $\Delta G$ for the dimer ${\rm G}\choose{\rm A}$, as a function of 
$T$. At $\Delta G_0=0$ and $s=0$, the melting temperature $T_{\rm m}$ of ${\rm G}\choose{\rm A}$, computed using 
$\Delta G(T_{\rm m})=0$, increases with $h$ and equals $T_{\rm m}$ in Table \plainref{tbl:table1} when $h=5.98$ kcal/mol. 
If $s=0$, the entropy loss of stacking, given by the slope of $\Delta G(T)$ over $T$, is smaller than the value of 
$\Delta S$ specified in Table \plainref{tbl:table1}. To rectify this discrepancy we take 
$U_{\rm ST}^0=-5.98+k_{\rm B}(T-T_{\rm m})s$ with $s>0$, which does not alter $T_{\rm m}$ but allows us to adjust the 
slope of $\Delta G(T)$ by adjusting the value of $s$. We find that $s=5.30$ is consistent with $\Delta S$ of 
${\rm G}\choose{\rm A}$ in Table \plainref{tbl:table1}. 

We carried out the same fitting procedure for all coarse-grained dimers. The resulting parameters $U_{\rm ST}^0$ are 
summarized in Table \plainref{tbl:table2} for $\Delta G_0=0.6$ kcal/mol ($\Delta G_0\approx k_{\rm B}T$ at room temperature). This 
value of $\Delta G_0$ gives the best agreement between simulation and experiment (see also Results and Discussion). Note 
that, although some stacks have equivalent thermodynamic parameters in Table \plainref{tbl:table1}, they have somewhat different 
$U_{\rm ST}^0$ due to their geometrical differences.

Finally, the parameters $U_{\rm ST}^0$ in Table \plainref{tbl:table2} are coupled to the specific choice of $k_r$ and 
$k_{\phi}$ in Eq.~(\plainref{UST}), since these coefficients determine how much entropy is lost upon formation of a model stack. For 
any reasonable choice of $k_r$ and $k_{\phi}$, the coarse-grained simulation model without explicit solvent will require
correction factors $s$ to match the experimental $\Delta S$. If different values 
of $k_r$ and $k_{\phi}$ are chosen, the accuracy of the model will not be compromised as long as $U_{\rm ST}^0$ ($h$ and $s$) are 
also readjusted following the fitting procedure outlined above.

\subsection {Hydrogen Bond Interactions}

To model the RNA structures shown in Figure \plainref{SS}, we use coarse-grained hydrogen bond interactions $U_{\rm HB}$ which mimic 
the atomistic hydrogen bonds present in the folded structure. The atomistic structures of hairpins L8 and L10 have not 
been determined experimentally. We assume that the only hydrogen bonds stabilizing these hairpins come from six 
Watson-Crick base pairs in the hairpin stem. The NMR structure for the MMTV PK is available (PDB code 1RNK~\cite{Shen95JMB}). 
For the MMTV PK, we generated an optimal network of hydrogen bonds by submitting the NMR structure to the WHAT IF server 
at {\ttfamily http://swift.cmbi.ru.nl}. Each hydrogen bond is modeled by a coarse-grained interaction potential,
\begin{eqnarray}
U_{\rm HB}=U_{\rm HB}^0\times\left[1+5(r-r_0)^2+1.5(\theta_1-\theta_{1,0})^2+1.5(\theta_2-\theta_{2,0})^2\right.\nonumber\\
\left.+0.15(\psi-\psi_{0})^2+0.15(\psi_1-\psi_{1,0})^2+0.15(\psi_2-\psi_{2,0})^2\right]^{-1},
\label{UHB}
\end{eqnarray}
where $r$, $\theta_1$, $\theta_2$, $\psi$, $\psi_1$ and $\psi_2$ are defined in Figure \plainref{fgr:HBONDS} for different 
coarse-grained sites. For Watson-Crick base pairs, the equilibrium values $r_0$, $\theta_{1,0}$, $\theta_{2,0}$, $\psi_0$, 
$\psi_{1,0}$ and $\psi_{2,0}$ are adopted from the coarse-grained structure of an ideal A-form RNA helix\cite{IdealRNAsite}. 
For all other bonds, the equilibrium parameters are obtained by coarse-graining the PDB structure of the RNA molecule. 
Our approach assumes that an A-form helix is an equilibrium state for RNA canonical secondary structure.
Modeling of non-canonical base pairing and of the tertiary interactions is biased to the native structure.
The coefficients 5, 1.5 and 0.15 in Eq.~(\plainref{UHB}) were determined from the same simulations as $k_r$ and 
$k_{\phi}$ in Eq.~(\plainref{UST}). Equation~(\plainref{UHB}) specifies $U_{\rm HB}$ for a single hydrogen bond and it must be 
multiplied by a factor of 2 or 3 if the same coarse-grained sites are connected by multiple bonds (as in base pairing). 
The geometry of $U_{\rm HB}$ in Eq.~(\plainref{UHB}) is the minimum necessary to maintain stable helices in the 
coarse-grained model. In particular, simulations of the MMTV PK (Figure \plainref{SS}) at 10 $^{\circ}$C yield the RMS deviation 
from the NMR structure of 1.4 and 2.0 \r{A} for stems 1 and 2, respectively.

In the present implementation of the model, the only hydrogen bonds included in simulation are those that are found in the
PDB structure of the RNA molecule. However, large RNA molecules may have alternative patterns of secondary structure
that are sufficiently stable to compete with the native fold. To account for this possibility, we have developed an extended 
version of the model where we allow the formation of any G$-$C, A$-$U or G$-$U base pair. Although easily 
implemented, this additional feature makes simulations significantly less efficient due to a large number of base pairing 
possibilities. A description of the extended model and its implementation for large RNA will be reported separately.
For small RNA molecules, similar to the ones considered here, we find that the folding thermodynamics is largely 
unaffected by the inclusion of alternative base pairing.

\subsection{Electrostatic interactions}

To model electrostatic interactions, we employ the Debye-H{\"u}ckel approximation combined with the concept of counterion 
condensation\cite{Manning69JCP}, which has been used previously to determine the reduced charge on the phosphate groups
in RNA\cite{Heilman01JMBa}. The highly negatively charged RNA attracts counterions, which condense onto the sugar-phosphate backbone. The loss 
in translational entropy of a bound ion (in the case of spherical counterions) is compensated by an effective binding 
energy between the ion and RNA, thus making counterion condensation favorable. Upon condensation of counterions onto 
the RNA molecule, the charge of each phosphate group decreases from $-e$ to $-Qe$, where $Q<1$ and $e$ is the proton charge. 
The uncondensed mobile ions are described by the linearized Poisson-Boltzmann (or Debye-H{\"u}ckel) equation. It can be 
shown that the electrostatic free energy of this system is given by~\cite{Sharp90JPC}
\begin{equation}
G_{\rm DH}=\frac{Q^2e^2}{2\varepsilon}\sum_{i,j}\frac{\exp\left(-|{\bf r}_i-{\bf r}_j|/\lambda\right)}{|{\bf r}_i-{\bf r}_j|},
\label{GDH}
\end{equation}
where  $|\mathbf{r}_i-\mathbf{r}_j|$ is the distance between two phosphates $i$ and $j$, $\varepsilon$ is the dielectric 
constant of water and $\lambda$ is the Debye-H{\"u}ckel screening length. The value of the Debye length $\lambda$ must be 
calculated individually for each buffer solution using
\begin{equation}
\lambda^{-2}=\frac{4\pi}{\varepsilon k_{\rm B}T}\sum_n q_n^2\rho_n,
\label{lambda}
\end{equation}
where $q_n$ is the charge of an ion of type $n$ and $\rho_n$ is its number density in the solution. If evaluated in units 
of \r{A}$^{-3}$, the number density $\rho$ is related to the molar concentration $c$ through $\rho = 6.022 \times10^{-4}c$.
In the simulation model, the free energy $G_{\rm DH}$ is viewed as the effective energy of electrostatic interactions 
between RNA phosphates, $U_{\rm EL}=G_{\rm DH}$, and as such it contributes an extra term to the energy function in 
Eq.~(\plainref{TIS}). This implicit inclusion of the ionic buffer significantly speeds up simulations, leading to much enhanced 
sampling of RNA conformations.

To complete our description of $U_{\rm EL}$, we still need to define the magnitude of the phosphate charge 
$Q$. For rod-like polyelectrolytes in monovalent salt solutions, Manning's theory of counterion condensation 
predicts\cite{Manning69JCP}
\begin{equation}
Q = Q^*(T)=\frac{b}{l_{{\rm B}}(T)},
\label{Manning}
\end{equation}
where $b$ is the length per unit (bare) charge in the polyelectrolyte and $l_{{\rm B}}$ is the Bjerrum length,
\begin{equation}
l_{{\rm B}}=\frac{e^2}{\varepsilon k_{\rm  B}T}.
\label{lB}
\end{equation}
According to Eq.~(\plainref{Manning}), the reduced charge $Q$ ($Q=1$ in the absence of counterion condensation) does not depend on 
the concentration $c$ of monovalent salt. The dependence of $Q$ on $T$ is nonlinear, since the dielectric constant of 
water decreases with the 
temperature\cite{Hasted72Book},
\begin{equation}
\varepsilon(T)=87.740-0.4008T+9.398\times10^{-4}T^2-1.410\times10^{-6}T^3,
\label{eps}
\end{equation}
where $T$ is in $^{\circ}$C. 

We estimate $b$ in Eq.~(\plainref{Manning}) from available folding data for hairpins L8 and L10, which were 
measured extensively in monovalent salt solutions of different ionic strength~\cite{Williams96Biochem}. We find that $b=4.4$ \r{A} 
reproduces measured stabilities of these hairpins over a wide range of salt concentrations. Assuming $b=4.4$ \r{A} for 
any RNA in a monovalent salt solution, we obtain good agreement between simulation and experiment for the MMTV PK 
(Figure \plainref{SS}). We propose that Eq.~(\plainref{GDH}) is sufficient to describe salt dependencies 
of RNA structural elements such as double helices, loops and pseudoknots.

In our coarse-grained simulation model, individual charges are placed at the centers of mass of the phosphate groups 
(sites P). This can be compared to an atomistic representation of the phosphate group, where the negative charge is 
concentrated on the two oxygen atoms. A more detailed distribution of the phosphate charge is not expected to have a 
significant effect on the electrostatic interactions between different strands in an RNA molecule. For instance, the 
distance between two closest phosphate groups on the opposite strands of a double helix is approximately 18 \r{A}, as 
compared to the distance between two atoms in a phosphate group of about 1 \r{A}. Furthermore, when considering a 
single strand, the dominant effect of the backbone charge distribution will be to modulate the magnitude of the reduced 
charge $Q$. If the density of the bare backbone charge is slightly underestimated then Eq.~(\plainref{Manning}) will predict a larger 
value of $Q$. Therefore, fitting $Q$ to experimental data allows us to compensate for small scale variations in the 
backbone charge density.

\subsection{Calculation of Stabilities}

We are interested in calculating the stability $\Delta G$ of the RNA structures shown in Figure \plainref{SS} as a function of 
temperature $T$. However, the folded and unfolded states of RNA coexist only in a narrow range of $T$ around the melting 
temperature. Thus, computing $\Delta G$ by means of direct sampling of the folding/unfolding transition at any $T$ is not 
feasible. Below we derive a formula for $\Delta G (T)$ from fundamental thermodynamic relationships that enables us to 
circumvent this problem.

Consider the Gibbs free energy of the folded state, $G_{\rm  f}=H_{\rm  f}-TS_{\rm  f}$. We can write the 
following exact expressions for the enthalpy $H_{\rm  f}$ and entropy $S_{\rm  f}$,
\begin{eqnarray}
H_{\rm  f}(T)=H_{\rm  f}(T^*)+\int_{T^*}^{T}\frac{\partial H_{\rm  f}}{\partial T}dT,\nonumber\\
S_{\rm  f}(T)=S_{\rm  f}(T^*)+\int_{T^*}^{T}\frac{\partial S_{\rm  f}}{\partial T}dT,
\label{HS}
\end{eqnarray}
where $T^*$ is an arbitrary reference temperature. The derivatives in Eq.~(\plainref{HS}) can be expressed in terms of the 
heat capacity $C_{\rm  f}$,
\begin{equation}
C_{\rm  f}=\frac{\partial H_{\rm  f}}{\partial T}=T\frac{\partial S_{\rm  f}}{\partial T},
\label{Cp}
\end{equation}
so that
\begin{equation}
G_{\rm  f}(T)=H_{\rm  f}(T^*)-TS_{\rm  f}(T^*)+\int_{T^*}^{T}C_{\rm  f}dT-T\int_{T^*}^{T}\frac{C_{\rm  f}}{T} dT.
\label{GT1}
\end{equation}
If we assume that the heat capacity of the folded state does not change significantly over the temperature range of 
interest, Eq.~(\plainref{GT1}) simplifies to
\begin{equation}
G_{\rm  f}(T)=H_{\rm  f}(T^*)-TS_{\rm  f}(T^*)-C_{\rm  f}\left(T^*-T+T\ln{\frac{T}{T^*}}\right).
\label{GT2}
\end{equation}
According to Eq.~(\plainref{GT2}), we can deduce the free energy $G_{\rm  f}$ of the folded state at temperature $T$ from the 
thermodynamic properties at some other temperature $T^*$. The same result holds for the free energy $G_{\rm  u}(T)$ 
of the unfolded state.

In the analysis of two-state transitions, it is convenient to use the transition (melting) temperature $T_{\rm  m}$ as 
the reference temperature for both folded and unfolded states. Then, the free energy difference $\Delta G$ between the 
folded and unfolded states is given by
\begin{equation}
\Delta G(T)=\Delta H(T_{\rm m})\left(1-\frac{T}{T_{\rm m}}\right)-\Delta C\left(T_{\rm m}-T+T\ln{\frac{T}{T_{\rm  m}}}\right),
\label{GT3}
\end{equation}
where we have used $\Delta G (T_{\rm m})=0$. Equation~(\plainref{GT3}) is commonly used to determine 
RNA stability from calorimetry experiments\cite{Mikulecky06Biopolym}, since it expresses $\Delta G(T)$ in terms of measured changes in enthalpy
and heat capacity. 

In simulations, we calculate the stability $\Delta G(T)$ of the folded RNA as follows. For each RNA illustrated in 
Figure \plainref{SS}, we run a series of Langevin dynamics simulations at different temperatures $T$ in the range from 0 to 
130 $^{\circ}$C. Using the weighted histogram technique, we combine the simulation data from all $T$ to obtain the 
density of energy states, $\rho (E)$, which is independent of temperature. The total free energy of the system, $G(T)$, 
is then given by
\begin{equation}
G(T)=-k_{\rm  B}T\ln\int\rho(E)\exp\left(-\frac{E}{k_{\rm  B}T}\right)dE,
\label{GT4}
\end{equation}
where the integral, representing the partition function, runs over all energy states.  At low $T$, the partition function 
in Eq.~(\plainref{GT4}) is dominated by the folded conformations and therefore, $G(T)\approx G_{\rm  f}(T)$. This allows us to rewrite 
Eq.~(\plainref{GT2}) as
\begin{equation}
G_{\rm  f}(T)=G(T^*)+\frac{\partial G}{\partial T}(T^*)\left(T-T^*\right)
+T\frac{\partial^2 G}{\partial T^2}(T^*)\left(T^*-T+T\ln{\frac{T}{T^*}}\right),
\label{GT5}
\end{equation}
where we take $T^*=0$ $^{\circ}$C to be the reference temperature for the folded state. To obtain Eq.~(\plainref{GT5}), we used
$S=-\partial G/\partial T$ and $C=-T\partial^2 G/\partial T^2$ . Equation~(\plainref{GT5}) can also be used to compute 
the free energy of the unfolded state, $G_{\rm  u}(T)$, if the reference temperature $T^*$ is chosen such that 
$G(T^*)\approx G_{\rm u}(T^*)$ --- for example, $T^*=130$ $^{\circ}$C. The stability $\Delta G(T)$ of the folded RNA is 
given by a difference between $G_{\rm  f}(T)$ and $G_{\rm u}(T)$, as illustrated in Figure \plainref{dGillust}. 

The present calculation is an alternative to the commonly used order parameter method to determine $\Delta G$ and can be 
applied to any folding/unfolding transition without further adjustments. Furthermore, in contrast to Eq.~(\plainref{GT3}), our 
approach will still work in systems that do not exhibit a two-state behavior, since it employs different reference 
temperatures for the folded and unfolded states. The only assumption is that at the reference temperature $T^*$ for the
folded (unfolded) state the population of the unfolded (folded) state is negligible. At the reference temperatures
chosen in our simulation, 0 $^{\circ}$C and 130 $^{\circ}$C, this assumption is trivially satisfied.

The formalism described above, including the weighted histogram technique, assumes that the conformational energy
$E$ in Eq.~(\plainref{GT4}) does not depend on temperature. However, the stacking interactions in Eq.~(\plainref{UST}) and electrostatic 
interactions in Eq.~(\plainref{GDH}) have $T$ as a parameter. The stacking parameters $U_{\rm ST}^0$ in Eq.~(\plainref{UST}) are 
linear in $T$, so we can write $U_{\rm ST}=u_0+k_{\rm  B}Tu_1$, where $u_0$ and $u_1$ are temperature independent. 
The Boltzmann factor of the second term, $\exp(-u_1)$, does not contain $T$ and cannot affect the temperature dependence of 
thermodynamic quantities. In the data analysis using weighted histograms, it is convenient to incorporate this Boltzmann 
factor into the density of states $\rho(E)$. Effectively, this means that the stacking interactions $u_1$ can be omitted 
from the total energy $E$ in Eq.~(\plainref{GT4}) and from all ensuing formulas. The electrostatic interactions in Eq.~(\plainref{GDH}) depend 
on $T$ nonlinearly, since $Q$, $\varepsilon$ and $\lambda$ are all functions of $T$. However, we operate within a 
relatively narrow range of temperatures --- the thermal energy $k_{\rm  B}T$ is between 0.54 and 0.8 kcal/mol. This 
justifies expanding the electrostatic potential $U_{\rm EL}$ up to the first order in $T$, which then enables us to 
treat it similarly to $U_{\rm ST}$. We expand $U_{\rm EL}$ around 55 $^{\circ}$C, in the middle of the relevant
temperature range. We have checked that this linear expansion does not affect the numerical results reported here.

\subsection {Langevin Dynamics Simulations}

The RNA dynamics are simulated by solving the Langevin equation, which for bead $i$ is 
$m_i\ddot{\mathbf{r}}_i=-\gamma_i\dot{\mathbf{r}}_i+\mathbf{F}_i+\mathbf{f}_i$, where $m_i$ is the bead mass, 
$\gamma_i$ is the drag coefficient, $\mathbf{F}_i$ is the conservative force, and $\mathbf{f}_i$ is the Gaussian random 
force, $\left<\mathbf{f}_i(t)\mathbf{f}_j(t^{\prime})\right>=6k_{\rm  B}T\gamma_i\delta_{ij}\delta(t-t^{\prime})$. 
The bead mass $m_i$ is equal to the total molecular weight of the chemical group associated with a given bead. The drag 
coefficient $\gamma_i$ is given by the Stokes formula, $\gamma_i=6\pi \eta R_i$, where $\eta$ is the viscosity of the 
medium and $R_i$ is the bead radius. To enhance conformational sampling~\cite{Honeycutt92Biopolym}, we take 
$\eta=10^{-5}$Pa$\cdot$s, which equals approximately 1\% of the viscosity of water. The values of $R_i$ are 2 \r{A} for 
phosphates, 2.9 \r{A} for sugars, 2.8 \r{A} for adenines, 3 \r{A} for guanines and 2.7 \r{A} for cytosines and uracils.  
The Langevin equation is integrated using the leap-frog algorithm with a time step $\Delta t=2.5$ fs. 

\section {Results and Discussion}

There are three adjustable parameters in the model: the corrective constant $\Delta G_0$ in Eq.~(\plainref{eqn:dGsim}),
the strength of hydrogen bonds $U_{\rm HB}^0$ in Eq.~(\plainref{UHB}), and the length $b$, which defines the reduced 
phosphate charge $Q$ in Eq.~(\plainref{Manning}). The absolute value of the correction $\Delta G_0$ should be relatively small, i.e., 
$|\Delta G_0|<1$ kcal/mol. If our approach is successful, various RNA structures will be characterized by similar values 
of $\Delta G_0$ and $U_{\rm HB}^0$. The physical meaning of the variable $Q$ implies that $Q<1$. The precise value of $Q$ 
may depend on the specific RNA structure as well as the buffer properties, since both could determine the extent of 
counterion condensation. However, we find that $Q$ does not vary much for different monovalent salt buffers or different 
RNA.

\subsection{Calibration of $\Delta G_0$ and $U_{\rm HB}^0$}

The parameters $\Delta G_0$ and $U_{\rm HB}^0$ were adjusted to match the differential scanning calorimetry melting 
curve (or heat capacity) of the MMTV PK~\cite{Theimer00RNA} at 1 M Na$^{+}$ (Figure \plainref{VPK_c1}). The list of all hydrogen bonds 
in the MMTV PK structure is given in \ref {HB-MMTV}. The secondary structure of the MMTV PK comprises five base 
pairs in stem 1 and six base pairs in stem 2 (Figure \plainref{SS}). The tertiary structure is limited to singular hydrogen bonds 
and is not stable in the absence of Mg$^{2+}$ ions.~\cite{Shen95JMB}

We find that in simulations with $c=1$ M in Eq.~(\plainref{lambda}), the thermodynamic properties obtained are not sensitive to the 
magnitude of the phosphate charge. In particular, the simulation model yields similar heat capacities for the bare 
phosphate charge, $Q=1$, or if we assume counterion condensation, $Q = Q^*(T)$. This is not unexpected,
since the electrostatic interactions are screened at high salt concentration and do not contribute significantly to the 
RNA stability. Therefore, we can identify $\Delta G_0$ and $U_{\rm HB}^0$ which are $Q$-independent.

The measured heat capacity at $c=1$ M is reproduced well in simulation with $\Delta G_0=0.6$ kcal/mol and 
$U_{\rm HB}^0=2.43$ kcal/mol (Figure \plainref{VPK_c1}). The model correctly describes the overall shape of the melting curve,
including two peaks that indicate the melting transitions of the two stems. Stem 1 in the MMTV PK is comprised entirely of G$-$C base pairs 
(Figure \plainref{SS}) and, despite being shorter than stem 2, melts at a higher temperature. Although the melting temperature of 
stem 2 is reproduced very accurately in simulation, the melting temperature of stem 1 is somewhat underestimated, 
89 $^{\circ}$C instead of 95 $^{\circ}$C. We speculate that the failure to precisely reproduce both peaks is due to 
inaccurate estimates of the stacking parameters $U_{\rm ST}^0$ at high temperatures. In addition to several approximations 
involved in the derivation of $U_{\rm ST}^0$, the experimental data that were used in this derivation were obtained at 
37 $^{\circ}$C or below. The approximation that enthalpies and entropies are constant may not be accurate for large 
temperature extrapolations. It is therefore expected that the agreement between experiment and simulation will be 
compromised at high temperatures. 

In the rest of the paper we set $\Delta G_0=0.6$ kcal/mol and $U_{\rm  {HB}}^0=2.43$ kcal/mol. Since the magnitude of 
the backbone charge could not be determined at high salt concentration, we will analyse the measurements of hairpin 
stability that cover a wide range of $c$. In this analysis we assume that $Q$ is given by Eq.~(\plainref{Manning}), 
with $b$ constant.

\subsection{Determination of $b$}

To estimate the reduced phosphate charge $Q$, we have computed the stabilities $\Delta G(c)$ of hairpins L8 and L10 
(Figure \plainref{SS}) for different values of $b$ in Eq.~(\plainref{Manning}). We use these hairpins as benchmarks because their folding 
enthalpies and entropies have been measured over a wide range of $c$, from 0.02 M to 1 M Na$^{+}$. The experimental 
$\Delta G(c)$ of L8 and L10 increase linearly with $\ln c$ for $c<0.11$ M, but the extrapolation of this linear dependence 
to $c>0.11$ M does not yield the measured stabilities at 1 M salt (Figure \plainref{L1}). In addition, the measured stability of L10 
at 1 M is disproportionately larger than that of L8. For these reasons, we use 0.11 M as a reference salt concentration 
$c_0$, instead of usual 1 M, and compare simulation and experiment in terms of relative stabilities 
$\Delta\Delta G(c)=\Delta G(c)-\Delta G(c_0)$. 

The simulation model reproduces correctly the linear dependence of $\Delta\Delta G$ on $\ln c$,
\begin{equation}
\Delta\Delta G(c) = -k_c\ln\frac{c}{c_0}
\label{ddG}
\end{equation}
for $c<0.11$ M. It also predicts an upward curvature of $\Delta\Delta G(c)$ for $c>0.11$ M (Figure \plainref{L1}). We find that 
$b=4.4$ \r{A} yields the best fit between the simulation and experimental values of $k_c$. Note that, although the 
stability of RNA hairpins decreases sharply with temperature, the salt dependence of $\Delta G$ is mostly insensitive to 
$T$ (Figure \plainref{L1}). The linear slope in Eq.~(\plainref{ddG}) does not change with temperature in experiment and simulation.

An uncertainty in the analysis of the L8 and L10 data comes from the 5'-pppG, which is subject to hydrolysis in solution.
The total number of phosphates $N_p$ may vary from 20 to 22 in L8 and from 22 to 24 in L10. Panels (c) and (d) in 
Figure \plainref{L1} show $\Delta\Delta G$ for the hairpins with charge $-Qe$, $-2Qe$ or $-3Qe$ at the 5'-end for 
$t=37$ $^{\circ}$C and $Q=0.60$ ($b=4.4$ \r{A}). Apparently, the charge of a terminal nucleotide has a strong influence 
on the hairpin stability. For DNA duplexes, the value of $k_c$ was shown to increase linearly with the total number of 
phosphates in the duplex,
\begin{equation}
k_c=0.057N_p.
\label{kc}
\end{equation}
This formula assumes implicitly that all phosphates contribute equally to the duplex stability. We find that for short RNA 
hairpins, such as L8 and L10, the end effects are significantly greater than $1/N_p$. In Figure \plainref{L1}, $k_c = 0.0445N_p$ for 
L8 with $N_p=22$ and $k_c = 0.0482N_p$ for L10 with $N_p=24$. Note that the ratio $k_c/N_p$ shifts towards its value in 
Eq.~(\plainref{kc}) with increasing the hairpin length. 

Although the experimental scatter in Figure \plainref{L1} can be attributed to partial hydrolysis of the 5'-pppG, it is hard to 
establish the precise contribution of this effect. Therefore, we fix $b=4.4$ \r{A}, which was obtained assuming no 
hydrolysis of the 5'-pppG.

At $c=0.11$ M, the simulation model predicts $T_{\rm m}=69.8$ $^{\circ}$C for the melting temperature of L8 and 
$\Delta G = -6.6$ kcal/mol for its stability at 37 $^{\circ}$C. The corresponding experimental values are 
$T_{\rm m}=75.7$ $^{\circ}$C and $\Delta G=-7.4$ kcal/mol at 37 $^{\circ}$C. For L10, we have $T_{\rm m}=66.5$ $^{\circ}$C, 
$\Delta G=-6.1$ kcal/mol in simulation and $T_{\rm m}=73.0$ $^{\circ}$C, $\Delta G=-6.6$ kcal/mol in experiment.
Both hairpins are found to be less stable in simulation than in experiment. Predictions of hairpin stability at 1 M salt, 
using the nearest neighbor model with stacking parameters from ref. 29, underestimate the melting temperatures 
and stabilities of L8 and L10 by a comparable amount. This suggests that some additional structuring may occur in the 
loops of these hairpins, which is not taken into account in theoretical models. Although our simulations account
for possible base stacking in the loops, we do not consider any hydrogen bonds other than the six Watson-Crick base pairs 
in the hairpin stem (Figure \plainref{SS}). It is possible that bases in the loops of L8 and L10 form additional hydrogen bonds,
since these loops are relatively large.

\subsection{Melting at low ionic concentration}

Figure \plainref{VPK_c005} compares the experimental heat capacity of the MMTV PK\cite{Theimer00RNA} at 50 mM K$^{+}$ to the result obtained 
in simulation with $c=0.05$ M in Eq.~(\plainref{lambda}). It is not obvious {\it a priori} that hairpins and pseudoknots should have 
the same reduced charge $Q$. Pseudoknot structures consist of three aligned strands of RNA, rather than two, and the high 
density of negative charge would be expected to promote counterion condensation. Nonetheless we find that the heat capacity
of the MMTV PK computed using $b=4.4$ \r{A}, which was established for hairpins, matches the experiment well 
(Figure \plainref{VPK_c005}b). Adjusting a single parameter $b$ was sufficient to position correctly both melting peaks, an indication 
that Eq.~(\plainref{GDH}) is suitable for a description of salt effects on RNA pseudoknots. The model also captures the characteristic 
property of the MMTV PK, that is, stem 2 is more strongly affected by changes in $c$ than stem 1. In experiment\cite{Theimer00RNA}, the 
difference in the melting temperatures of the two stems increases from 22 $^{\circ}$C at $c=1$ M to 32 $^{\circ}$C at 
$c=0.05$ M, which is related to a significant loss of stability for stem 2 in the low salt buffer. Note that neglecting 
counterion condensation ($Q=1$) overestimates electrostatic repulsions between phosphates, rendering both stems 
significantly less stable in simulation than in experiment (Figure \plainref{VPK_c005}a). In particular, the melting transition of 
stem 2 shifts to 4 $^{\circ}$C, in stark contrast to the experimental melting temperature of 48 $^{\circ}$C.

A considerable difference in the stability of the two stems in the MMTV PK is further illustrated in Figure \plainref{VPK_frac_fold}, 
where we plot the probability that each stem is folded as a function of $c$. At 37 $^{\circ}$C, stem 1 is stable for all
salt concentrations in the typical experimental range, whereas stem 2 undergoes a folding transition upon increasing $c$,
with the midpoint at approximately 30 mM (Figure \plainref{VPK_frac_fold}a). At 80 $^{\circ}$C, the folding transition of stem 1 falls 
within the experimental range of $c$ (Figure \plainref{VPK_frac_fold}b). However, at such high temperatures, the population of the 
unfolded state is non-negligible for all salt concentrations and, in the case of stem 2, it exceeds 80\%. 

In Figure \plainref{VPK_frac_fold}, we have used two different criteria for folding of the stems. For the solid curves a stem is 
considered folded if at least five base pairs have formed and for the dashed curves a stem is assumed to be folded if at least
one base pair has formed. Although the curves in Figure \plainref{VPK_frac_fold} depend on the criteria for folding, the numerical 
differences are small, especially at 37 $^{\circ}$C. This is because the transition state in the folding of each 
stem corresponds to the closing of a loop by a single base pair, after which the formation of subsequent base pairs is a 
highly cooperative process. At 80 $^{\circ}$C, individual base pairs have a high probability of opening and closing 
without affecting the loop region, which contributes to the quantitative differences between the two definitions 
of the folded state (Figure \plainref{VPK_frac_fold}b).

\section{Conclusions}

We have developed a general coarse-grained simulation model that reproduces the folding thermodynamics of RNA hairpins and 
pseudoknots with good accuracy. The model enables us to study the folding/unfolding  transitions with computational 
efficiency, as a function of temperature and ionic strength of the buffer. It is interesting that simulations using a 
single choice of model parameters, $\Delta G_0=0.6$ kcal/mol, $U_{\rm  {HB}}^0=2.43$ kcal/mol and $b=4.4$ \r{A}, show 
detailed agreement with available experimental data for the three RNA molecules in monovalent salt buffers. Although we 
have established the success of the model with applications to a few RNA molecules, the methodology is general and we 
expect that the proposed force field can be used to study RNA with even more complex structures. Applications of the model 
to other RNA molecules will be reported in a separate publication.

On the basis of the good agreement between simulations and experiments we conclude that, for $c<0.2$ M, the effects of 
monovalent salt on RNA stability can be attributed to the polyelectrolyte effect. At $c>0.2$ M, the results are more 
ambiguous both in experiment and simulation. There is mixed experimental evidence as to whether the linear dependence of 
the RNA stability on $\ln c$ extends all the way to 1 M (cf. L8 and L10 in Figure \plainref{L1}). Our simulations predict a 
substantial curvature in $\Delta\Delta G$ vs. $\ln c$ in the range $c>0.2$ M (Figure \plainref{L1}), where the Debye-H{\"u}ckel 
approximation is likely to be less accurate. However, the melting profile of the MMTV PK obtained in simulations at 1 M is 
in good agreement with experiment (Figure \plainref{VPK_c1}). Due to insufficient experimental data, it is hard to establish the 
extent to which simulations and experiments disagree at $c>0.2$ M. 

We find that, both for the hairpins and pseudoknots in monovalent salt solutions, the reduction in the magnitude of the 
backbone charge due to counterion condensation is given by Eq.~(\plainref{Manning}) with $b=4.4$ \r{A}. This result is particularly interesting since folded 
pseudoknots have a higher density of backbone packing than folded hairpins. In counterion condensation theory of rod-like 
polyelectrolytes, the parameter $b$ is the mean axial distance per unit bare charge of the polyelectrolyte.
Notably, distance 4.4 \r{A} agrees well with the estimates of the counterion condensation theory for $b$ in 
single-stranded nucleic acids~\cite{Manning76Biopolym,Record76Biopolym,Record94BJ}. Therefore, we propose that, in our 
simulations, $b$ describes the geometry of the unfolded state, which is similarly flexible for hairpins and pseudoknots. 
Further work on this issue and the reduction of RNA charge in the presence of divalent counterions will be published 
elsewhere. 


\providecommand*\mcitethebibliography{\thebibliography}
\csname @ifundefined\endcsname{endmcitethebibliography}
  {\let\endmcitethebibliography\endthebibliography}{}

\clearpage

\begin{table}
\renewcommand{\baselinestretch}{2}
\renewcommand{\arraystretch}{2}
\caption{Enthalpies $\Delta H$, entropies $\Delta S$ and melting temperatures $T_{\rm m}$ of single-stranded stacks, derived in 
this work. Enthalpies of hydrogen bond formation in Watson-Crick base pairs are given in the last two rows. In the first 
column, the $5'$ to $3'$ direction is shown by an arrow. }
\bigskip
\bigskip
\begin{center}
\begin{tabular}{c|c|c|c}
\hline\hline
$\uparrow\hspace{-1mm}{x\atop w}$ &  $\Delta H$, kcal$\,$mol$^{-1}$ & $\Delta S$, cal$\,$mol$^{-1}$K$^{-1}$ & $T_{\rm m}$,$^{\circ}$C \\ 
\hline
${\rm U}\atop{\rm U}$ & $-1.81$ & $-7.2$ & $-21$ \\
${\rm C}\atop{\rm C}$ & $-2.87$ & $-10.0$ & 13 \\
${\rm C}\atop{\rm U}$; ${\rm U}\atop{\rm C}$  & $-2.87$ & $-10.0$ & 13 \\ 
${\rm A}\atop{\rm A}$ & $-3.53$ & $-11.8$ & 26\\
${\rm A}\atop{\rm U}$; ${\rm U}\atop{\rm A}$ & $-3.53$ & $-11.8$ & 26 \\
${\rm A}\atop{\rm C}$; ${\rm C}\atop{\rm A}$ & $-3.53$ & $-11.8$ & 26\\ 
${\rm G}\atop{\rm C}$ & $-4.21$ & $-13.3$ & 42\\
${\rm G}\atop{\rm U}$; ${\rm U}\atop{\rm G}$ & $-5.55$ & $-16.4$ & 65 \\
${\rm C}\atop{\rm G}$ & $-6.33$ & $-18.4$ & 70\\ 
${\rm G}\atop{\rm A}$; ${\rm A}\atop{\rm G}$ & $-6.75$ & $-19.8$ & 68 \\
${\rm G}\atop{\rm G}$ & $-8.31$ & $-22.7$ & 93 \\
\hline
\multicolumn{4}{c}{$\Delta H({\rm A}-{\rm U})=-1.47$ kcal/mol }\\
\multicolumn{4}{c}{$\Delta H({\rm G}-{\rm C})=-2.21$ kcal/mol }\\
\hline\hline 
\end{tabular}
\end{center}
\label{tbl:table1}
\end{table}

\clearpage

\begin{table}
\renewcommand{\baselinestretch}{2}
\renewcommand{\arraystretch}{2}
\caption{Temperature-dependent stacking parameters $U_{\rm {ST}}^0$ used in Eq.~(\plainref{UST}). The values of $h$ correspond to 
$\Delta G_0=0.6$ kcal/mol in Eq.~(\plainref{eqn:dGsim}). The melting temperatures $T_{\rm m}$ are given in Table \plainref{tbl:table1}. 
In the first column, the $5'$ to $3'$ direction is shown by an arrow.}
\bigskip
\bigskip
\begin{center}
\begin{tabular}{c|c|c}
\hline\hline
& \multicolumn{2}{|c}{$U_{\rm  {ST}}^0=-h+k_{\rm B}(T-T_{\rm m})s$}\\
\hline
$\uparrow\hspace{-1mm}{x\atop w}$ &  $h$, kcal$\,$mol$^{-1}$ & $s$ \\ 
\hline
${\rm U}\atop{\rm U}$ & 3.37 & $-3.56$\\
${\rm C}\atop{\rm C}$ & 4.01 & $-1.57$ \\
${\rm C}\atop{\rm U}$; ${\rm U}\atop{\rm C}$ & 3.99; 3.99 & $-1.57$; $-1.57$\\ 
${\rm A}\atop{\rm A}$ & 4.35 & $-0.32$\\
${\rm A}\atop{\rm U}$; ${\rm U}\atop{\rm A}$ & 4.29; 4.31 & $-0.32$; $-0.32$\\
${\rm A}\atop{\rm C}$; ${\rm C}\atop{\rm A}$ & 4.29; 4.31 & $-0.32$; $-0.32$\\ 
${\rm G}\atop{\rm C}$ & 4.60 & 0.77\\ 
${\rm G}\atop{\rm U}$; ${\rm U}\atop{\rm G}$ & 5.03; 4.98 & 2.92; 2.92\\
${\rm C}\atop{\rm G}$ & 5.07 & 4.37\\
${\rm G}\atop{\rm A}$; ${\rm A}\atop{\rm G}$ & 5.12; 5.08 & 5.30; 5.30\\
${\rm G}\atop{\rm G}$ & 5.56 & 7.35\\
\hline\hline 
\end{tabular}
\end{center}
\label{tbl:table2}
\end{table}

\clearpage

\begin{table}
\renewcommand{\arraystretch}{2}
\caption{Hydrogen bonds in the MMTV PK}
\begin{center}
\begin{tabular}{c|c}
\hline\hline
Residues in contact &  Hydrogen bonds \\ 
\hline
G1-C19 & N1-N3; N2-O2; O6-N4\\
G2-C18 & N1-N3; N2-O2; O6-N4\\
C3-G17 & N4-O6; N3-N1; O2-N2\\
G4-C16 & N1-N3; N2-O2; O6-N4\\
G4-A26 & O2'-N6\\
G4-A27 & N2-N1\\
C5-G15 & N4-O6; N3-N1; O2-N2\\
C5-A27 & O2'-O2'\\
G7-G9 & O2'-OP2\\
U8-A33 & N3-N1; O4-N6\\
G9-C32 & N1-N3; N2-O2; O6-N4\\
G10-C31 & N1-N3; N2-O2; O6-N4\\
G11-C30 & N1-N3; N2-O2; O6-N4\\
C12-G29 & N4-O6; N3-N1; O2-N2\\
U13-G28 & N3-O6; O2-N1\\
G17-A27 & N3-N6; O2'-N6\\
C18-A25 & O2-N6\\
C19-A24 & OP1-N6\\
\hline\hline 
\end{tabular}
\end{center}
\label{HB-MMTV}
\end{table}

\clearpage

\section{Figures}

\begin{figure}
\begin{center}
\includegraphics[width=10.0cm,clip]{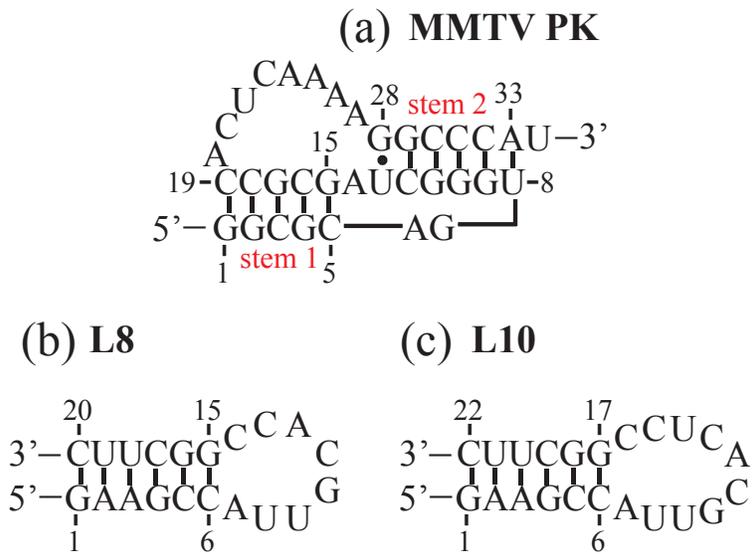}
\end{center}
\renewcommand{\baselinestretch}{2}
\caption{Secondary structures of studied RNA. Hairpins L8 and L10 have a 5'-pppG, while the MMTV PK does not have a 
phosphate group at the 5'-end.}
\label{SS}
\end{figure}

\clearpage

\begin{figure}
\begin{center}
\includegraphics[width=4.0cm,clip]{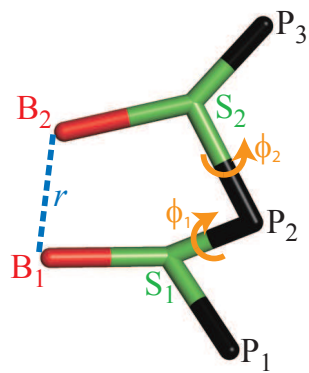}
\end{center}
\renewcommand{\baselinestretch}{2}
\caption{Illustration of the structural parameters in Eq.~(\plainref{UST}). Sites P, S and B are shown in black, green and red, 
respectively. The indices refer to different nucleotides. $r$ is the distance between bases ${\rm B}_1$ and ${\rm B}_2$
in angstroms, and $\phi_1 ({\rm P}_1, {\rm S}_1, {\rm P}_2, {\rm S}_2)$ and 
$\phi_2 ({\rm P}_3, {\rm S}_2, {\rm P}_2, {\rm S}_1)$ are the dihedral angles in radians.
}
\label{STACK}
\end{figure}

\clearpage

\begin{figure}
\begin{center}
\includegraphics[width=12.0cm,clip]{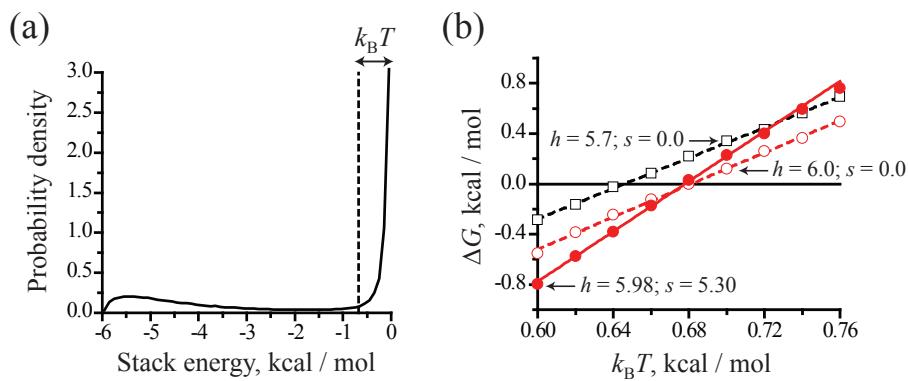}
\end{center}
\renewcommand{\baselinestretch}{2}
\caption{ (a) A sample distribution of stacking energies $U_{\rm  ST}$ (Eq.~(\plainref{UST})) from simulations 
of coarse-grained dimers. All dimer configurations with $U_{\rm ST}<-k_{\rm  B}T$ are counted as
stacked in Eq.~(\plainref{eqn:dGsim}). (b) The free energy, $\Delta G$, of stack formation in dimer ${\rm G}\choose{\rm A}$, 
calculated using Eq.~(\plainref{eqn:dGsim}) with $\Delta G_0=0$. Open and closed symbols show $\Delta G$ for different 
$U_{\rm  ST}^0=-h+k_{\rm  B}(T-T_{\rm  m})s$, where $T_{\rm  m}$ is the melting temperature of 
${\rm G}\choose{\rm A}$ in Table \plainref{tbl:table1} and $h$, $s$ vary. Red solid line shows $\Delta G=\Delta H-T\Delta S$ 
for $\Delta H$ and $\Delta S$ given in Table \plainref{tbl:table1}. Same $\Delta G$ is obtained in simulation with $h=5.98$ kcal/mol 
and $s=5.30$ (closed symbols).}
\label{fgr:UST1}
\end{figure}

\clearpage

\begin{figure}
\begin{center}
\includegraphics[width=12.0cm,clip]{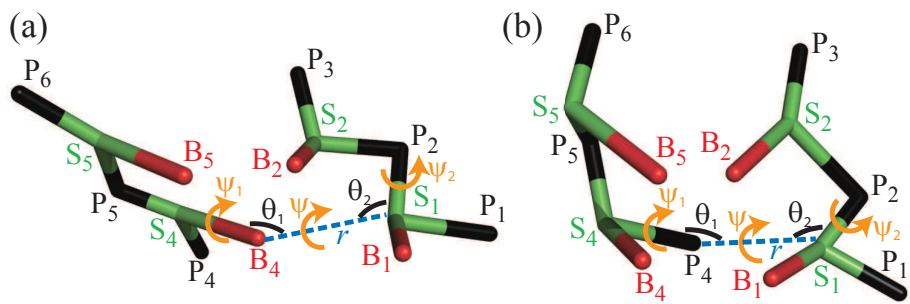}
\end{center}
\renewcommand{\baselinestretch}{2}
\caption{Illustration of the structural parameters in Eq. (\plainref{UHB}). Angle definitions depend on the site
(P, S or B) which forms a hydrogen bond. Examples show hydrogen bonding between sites B and S (a) and between sites P and S (b). 
In (a), $r$ (S1, B4) is distance, $\theta_1$ (S4, B4, S1) and $\theta_2$ (P2, S1, B4) are angles, 
$\psi$ (P2, S1, B4, S4), $\psi_1$ (S1, B4, S4, P5) and $\psi_2$ (B4, S1, P2, S2) are dihedral angles. 
In (b), $r$ (S1, P4) is distance, $\theta_1$ (S4, P4, S1) and $\theta_2$ (P2, S1, P4) are angles, 
$\psi$ (P2, S1, P4, S4), $\psi_1$ (S1, P4, S4, P5) and $\psi_2$ (P4, S1, P2, S2) are dihedral angles. All other 
designations are the same as in Figure \plainref{STACK}.}
\label{fgr:HBONDS}
\end{figure}

\clearpage

\begin{figure}
\begin{center}
\includegraphics[width=6.0cm,clip]{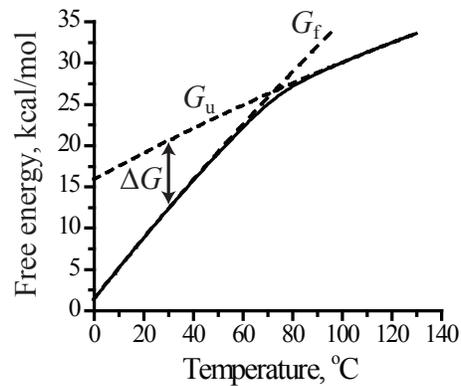}
\end{center}
\renewcommand{\baselinestretch}{2}
\caption{Geometrical definition of the stability $\Delta G(T)$ of the folded state. The solid curve shows the total free 
energy of the system $G(T)$, given by Eq.~(\plainref{GT4}). The free energies of the folded and unfolded states, $G_{\rm f}(T)$ 
and $G_{\rm u}(T)$, are estimated from Eq.~(\plainref{GT5}) using $T^*=0$ $^{\circ}$C and 130 $^{\circ}$C, respectively 
(dashed curves).}
\label{dGillust}
\end{figure}

\clearpage

\begin{figure}
\begin{center}
\includegraphics[width=12.0cm,clip]{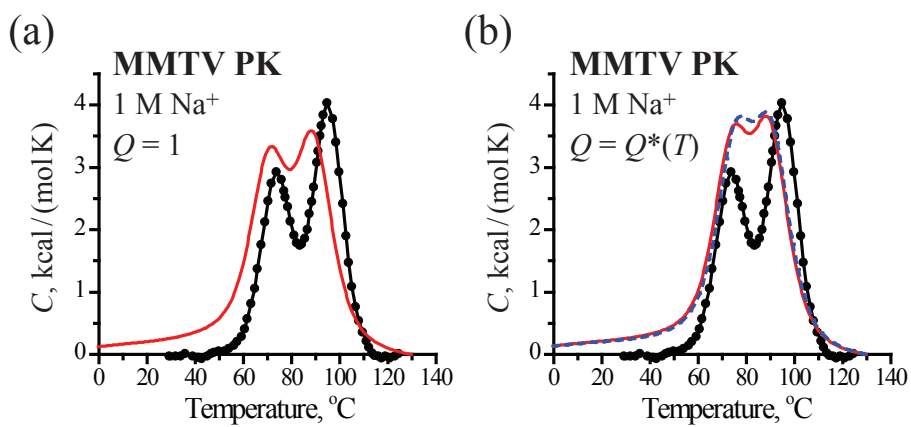}
\end{center}
\renewcommand{\baselinestretch}{2}
\caption{Measured\cite{Theimer00RNA} (black symbols) and computed (red curve) heat capacity $C$ of the MMTV PK in 1 M 
Na$^{+}$. The computed $C(T)$ is shown with respect to the heat capacity of the unfolded state at 130 $^{\circ}$C.
$\Delta G_0=0.6$ kcal/mol, $U_{\rm  {HB}}^0=2.43$ kcal/mol. In (a), $Q=1$. In (b), $Q=Q^*(T)$, $b=4.4$ \r{A}.
The blue dashed curve in (b) shows $C(T)$ from simulation using the solvent viscosity which is ten times larger than the
$\eta$ specified in Methods. It can be rigorously established that  the thermodynamic properties must be independent of $\eta$. In accord with this
expectation simulations at high and low values of $\eta$ are in good agreement with each other despite the difficulty in obtaining 
 adequate sampling for large $\eta$.}
\label{VPK_c1}
\end{figure}

\clearpage

\begin{figure}
\begin{center}
\includegraphics[width=12.0cm,clip]{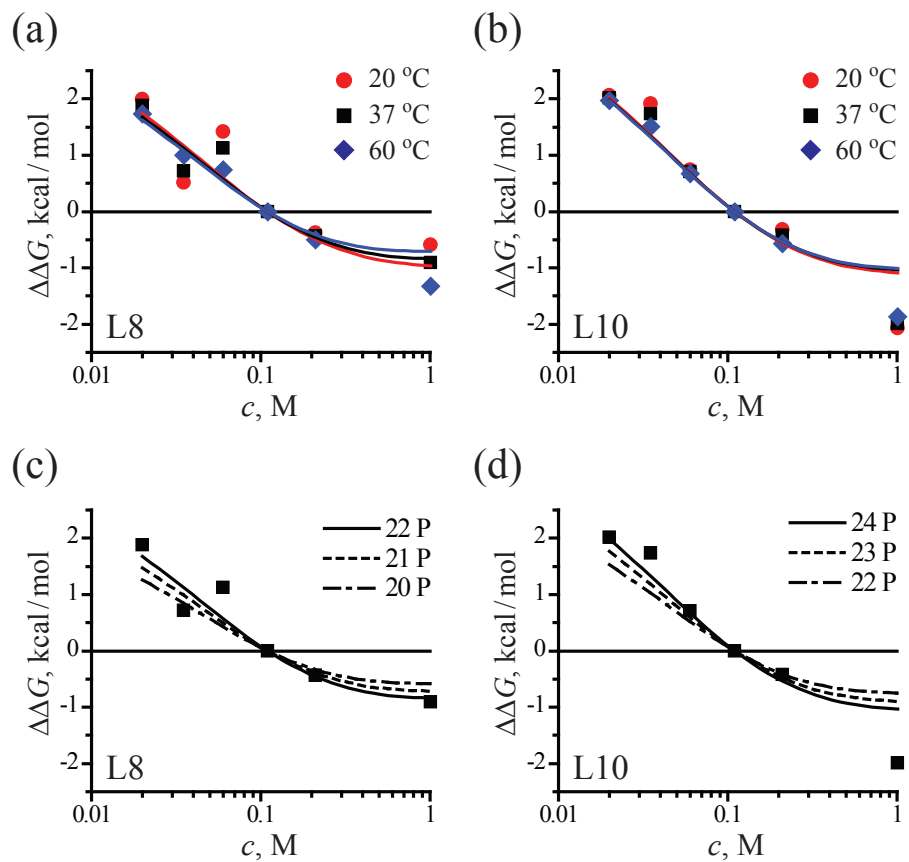}
\end{center}
\renewcommand{\baselinestretch}{2}
\caption{The stability $\Delta G$ of hairpins L8 and L10 as a function of salt concentration $c$, plotted as
$\Delta\Delta G(c)=\Delta G(c)-\Delta G(c_0)$, where $c_0=0.11$ M. Panels (a) and (b) show comparison of 
experiment (symbols) and simulation (curves) at different temperatures, assuming no hydrolysis of the 5'-pppG. 
Panels (c) and (d) illustrate the contribution of the 5'-pppG to $\Delta\Delta G(c)$ at 37 $^{\circ}$C for different 
levels of hydrolysis: no hydrolysis (solid), partial hydrolysis (dashed) and complete hydrolysis (dash-dotted). 
Simulation curves are for $Q=Q^*(T)$, $b=4.4$ \r{A}.}
\label{L1}
\end{figure}

\clearpage

\begin{figure}
\begin{center}
\includegraphics[width=12.0cm,clip]{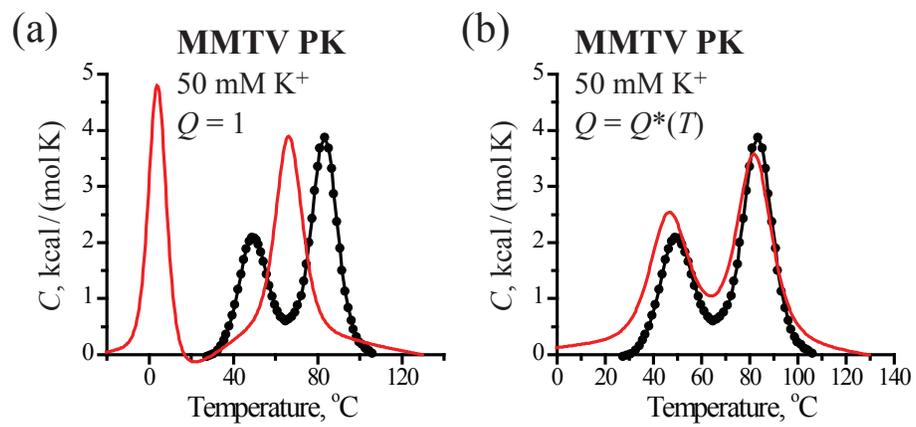}
\end{center}
\renewcommand{\baselinestretch}{2}
\caption{Same as in Figure \plainref{VPK_c1} but in 50 mM K$^{+}$.}
\label{VPK_c005}
\end{figure}

\clearpage

\begin{figure}
\begin{center}
\includegraphics[width=12.0cm,clip]{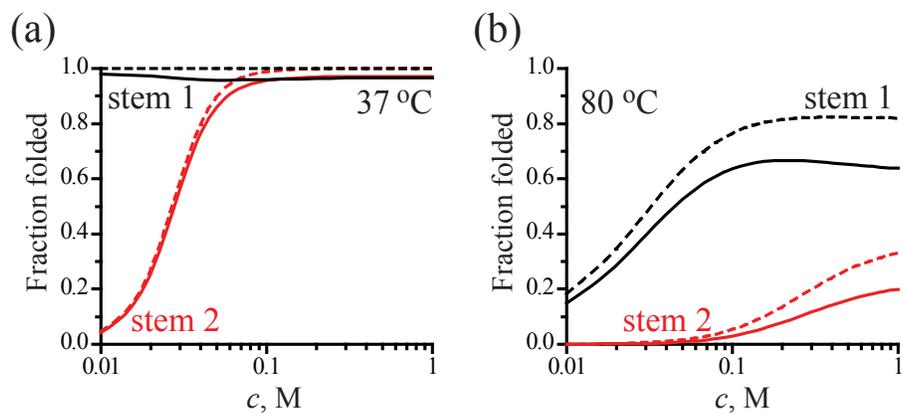}
\end{center}
\renewcommand{\baselinestretch}{2}
\caption{The fraction of folded stem 1 (black) and stem 2 (red) in the MMTV PK, as a function of salt concentration $c$. 
A stem is  folded if five base pairs have formed (solid lines) or if one base pair has formed (dashed 
lines). $\Delta G_0=0.6$ kcal/mol, $U_{\rm  {HB}}^0=2.43$ kcal/mol, $Q=Q^*(T)$, $b=4.4$ \r{A}. In (a), $T=37$ $^{\circ}$C.
In (b), $T=80$ $^{\circ}$C.}
\label{VPK_frac_fold}
\end{figure}






\end {document}